\documentclass[
    ,final            
  ]
  {aipproc}

\usepackage{times}
\usepackage{graphicx,bm,amssymb}
\usepackage{epsfig}
\usepackage{natbib}

\newcommand{\msun}{\mbox{$M_\odot$}}
\newcommand{\rsun}{\mbox{$R_\odot$}}
\def\be{\begin{eqnarray}}
\def\ee{\end{eqnarray}}
\def\lsim{\mathrel{\rlap{\lower3pt\hbox{\hskip1pt$\sim$}}
     \raise1pt\hbox{$<$}}} 
\def\gsim{\mathrel{\rlap{\lower3pt\hbox{\hskip1pt$\sim$}}
     \raise1pt\hbox{$>$}}} 

\layoutstyle{8x11single}

\begin{document}

\title{GRB/HN from Kerr Black Holes in Binaries}

\classification{97.80.Fk --- 98.70.Rz --- 97.60.Bw --- 97.60.Lf --- 97.80.Jp}
\keywords      {binaries: close --- gamma rays: bursts --- black hole physics --- supernovae: general --- X-rays: binaries}

\author{Enrique Moreno M\'endez}{
  address={{E-mail: emoreno@astro.uni-bonn.de} 
Argelander-Institut f\"ur Astronomie, Bonn University, Auf dem H\"ugel 71, 53121 Bonn, Germany}
}

\begin{abstract}
The Collapsar model, in which a massive star ($\gtrsim20\msun$) fails to produce a SN and forms a BH, provides the main framework for understanding long Gamma-Ray Bursts (GRB) and the accompanying hypernovae (HN).  However, single massive-star models that explain the population of pulsars, predict cores that rotate too slowly to produce GRBs/HNe.  We present a model of binary evolution that allows the formation of Kerr black holes (BH) where the spin of the BH can be estimated from the pre-collapse orbit, and use the Blandford-Znajek (BZ) mechanism to estimate the available energy for a GRB/HN.

A population synthesis study shows that this model can account for both, the long GRB and the subluminous GRB populations.

\end{abstract}

\maketitle

\section{Summary}

\begin{figure}
\centering
\begin{tabular}{ccc}
\includegraphics[height=45mm,width=0.48\textwidth]{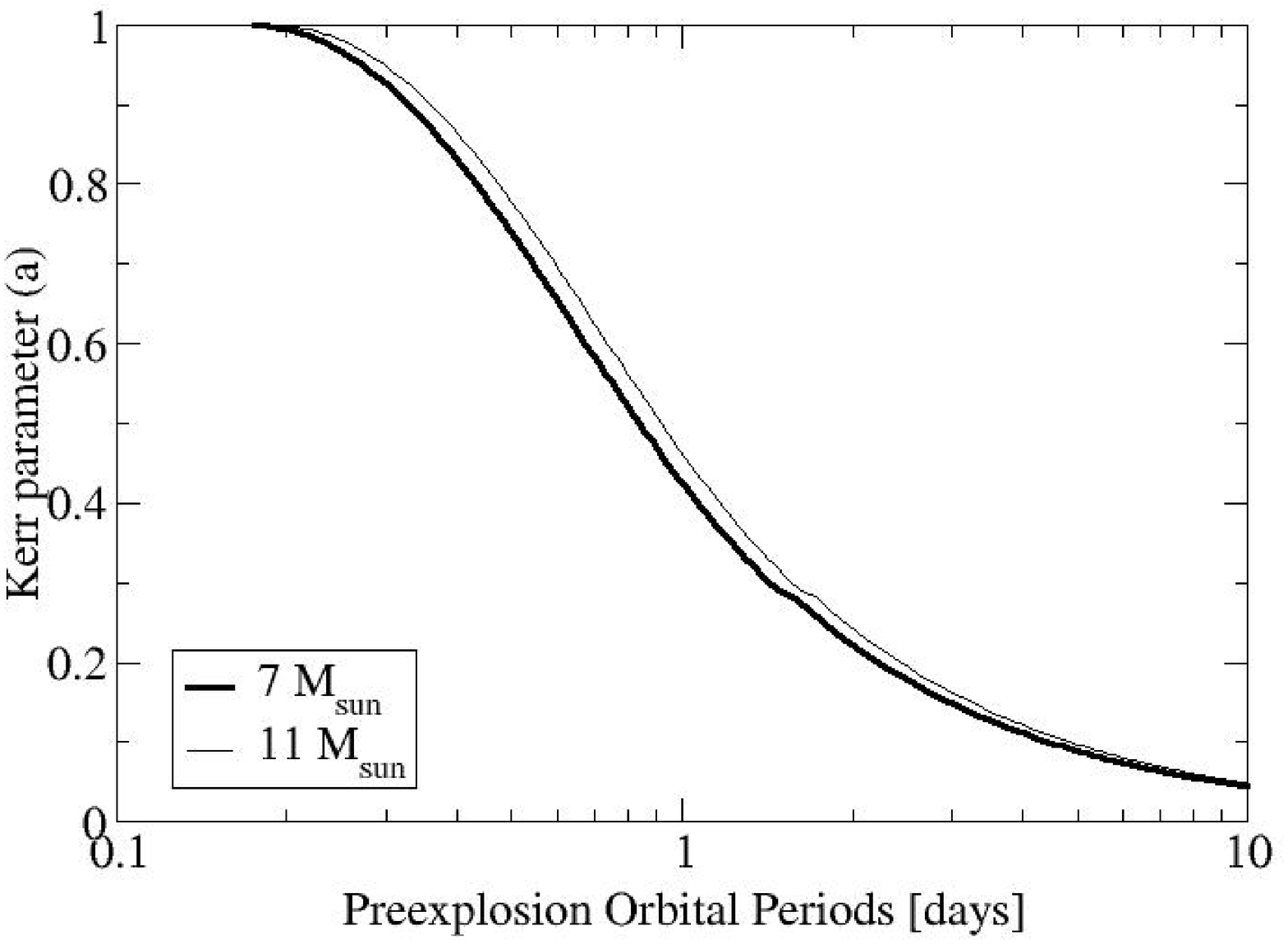}
& \hspace{2mm} &
\includegraphics[height=45mm,width=0.45\textwidth]{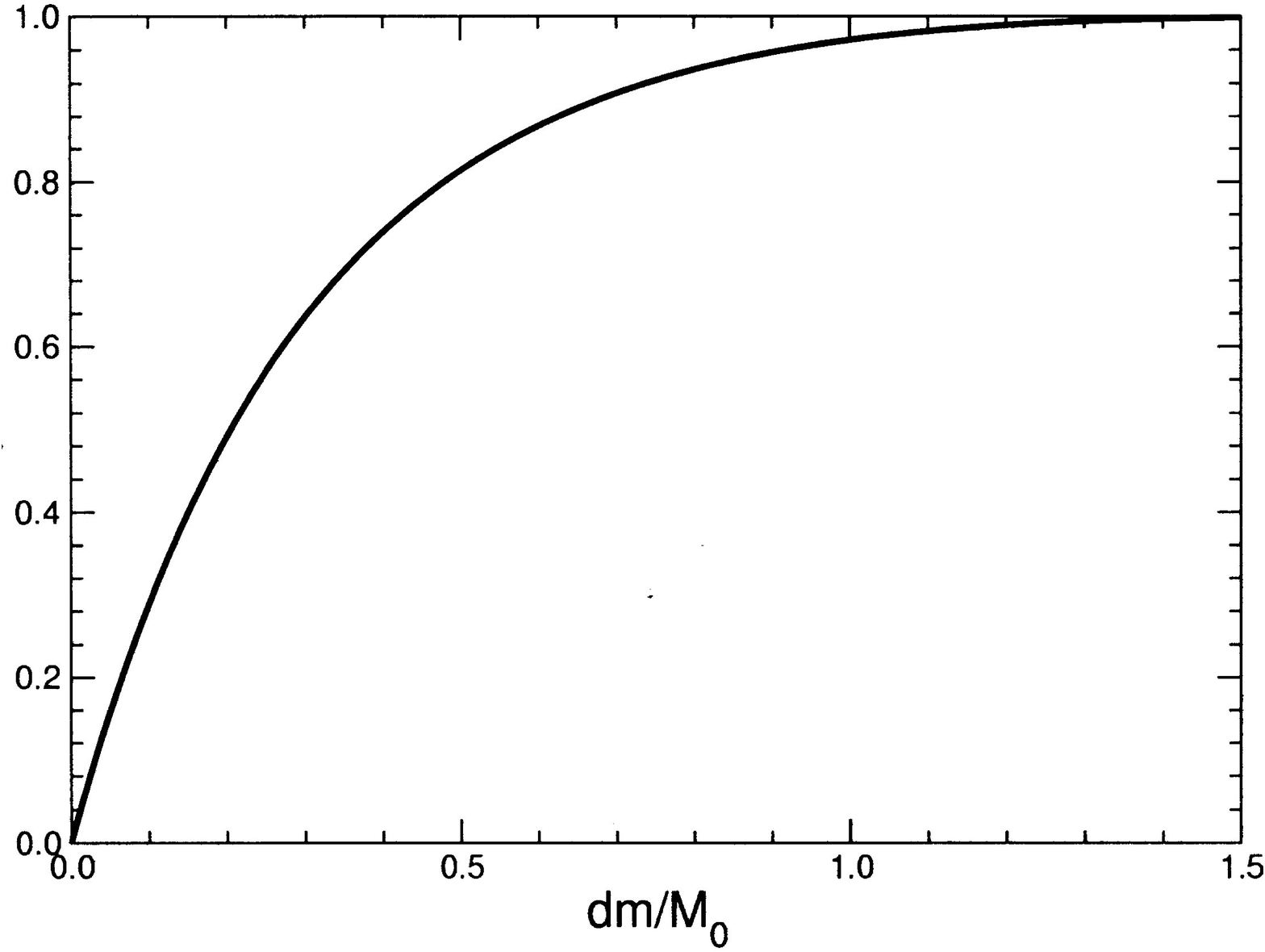}
\end{tabular}
\caption{\small{(Left) The BH spin $a_\star$ of the BH resulting from the collapse of a He star rotating synchronously with the orbit, as a function of orbital period \cite{Lee02}. The result depends very little on the mass of the He star, or on whether we use a simple polytrope or a more sophisticated model.
(Right) Spinning up BHs. $a_\star$ is given in units of [$GM/c^2$] and $\delta m$ is the total rest mass of the accreted material. Note that $M_0$ is the mass of the non-rotating initial BH; we assumed that the last stable orbit corresponds to the marginally stable radius \cite{Bro00}.}} \label{fig:1}
\end{figure}

We (\cite{Bro07}, \cite{BLMM08}, \cite{Lee02}, \cite{Mor07}, \cite{Mor08}) model binaries in which the primary star (BH progenitor) evolves as a single star until its core begins He burning.  When He burning begins in the core the stellar radius expands a further $10$ to $15\%$ relative to the red giant radius, it fills its Roche Lobe and begins to transfer mass to the secondary star, i.e. Case C mass transfer.  As a result, the orbit shrinks and the binary goes into a common envelope.  The H envelope of the primary is removed at the expense of orbital angular momentum.  Eventually the orbit, originally $>1,500\rsun$, decays to a few $\rsun$.  The ratio of the stellar radius to orbital separation leads the primary to spin up by tidal synchronization.  Hence, having a good estimate of the orbital period and, assuming angular momentum conservation during the collapse, one may obtain the natal Kerr parameter of the BH (see column 6 in Tab.~\ref{tab-results} and the left (L) plot in Fig.~\ref{fig:1}).

The natal Kerr parameter may be modified by two post-formation mechanisms.  The first one is by using the rotational energy to produce the GRB/HN, e.g. through the BZ mechanism \citep{Bla77}, a GRB/HN explosion (see column 9 in~\ref{tab-results}).  A drastic decrease of the natal Kerr parameter is not expected given that a small percentage of the available energy is enough to power the GRB/HN.  Further constraints on the energy of the explosion can be placed through the Blaauw-Boersma kick (\cite{Bla61}, \cite{Boe61}) and the fact that the binary remains bound (and, usually, has small eccentricity).  The second mechanism is by accreting mass into the BH (see the right (R) plot in Fig.~\ref{fig:1}) once the secondary, or its wind, fills its Roche Lobe.

Measurements by \cite{Sha06} coincide with the predictions of \cite{Lee02} for the Kerr parameters of GRO J1655$-$40 and 4U 1543$-$47.  Our estimates are consistent with the observed Kerr parameter in LMC X$-$3 (fig.3 in \cite{Davis06}), and the measurements for the Galactic GRS 1915$+$105 \citep{McC06}, as well as the extragalactic LMC X$-$1 \citep{Gou09} and M33 X$-$7 \citep{Liu08} can be explained through a phase of mass transfer (with wind mass transfer and hypercritical accretion for the last 2, see \cite{Mor08} and \cite{Mor10}).

\tiny{
\begin{table}
\centering
\begin{tabular}{|c|c|c|c|c|c|c|c|c|}
\hline
      Name     & $M_{BH,2}$ & $M_{d,2}$ &  $M_{BH,now}$ & $M_{d,now}$ &     Model     &    Measured   & $P_{Orbit,now}$ &   $E_{\rm BZ}$   \\
               & [$\msun$]  & [$\msun$] &   [$\msun$]   &  [$\msun$]  & $a_{\star,2}$ & $a_{\star}$ &      [days]     & [$10^{51}$ ergs]   \\
\hline
\hline
\multicolumn{9}{|c|}{AML: with main sequence companion} \\
\hline
J1118$+$480    &   $\sim5$  &   $<1$    &    $6.0-7.7$   &   $0.09-0.5$ &    $0.8$   &       -       &  $0.169930(4)$ &   $\sim 430$      \\
Vel 93         &   $\sim5$  &   $<1$    &    $3.64-4.74$ &  $0.50-0.65$ &    $0.8$   &       -       &     $0.2852$   &   $\sim 430$      \\
J0422$+$32     &    $6-7$   &   $<1$    &    $3.4-14.0$  &  $0.10-0.97$ &    $0.8$   &       -       &   $0.2127(7)$  & $500\sim 600$     \\
1859$+$226     &    $6-7$   &   $<1$    &     $7.6-12$   &              &    $0.8$   &       -       &    $0.380(3)$  & $500\sim 600$     \\
GS1124$-$683   &    $6-7$   &   $<1$    &     $6.95(6)$  &  $0.56-0.90$ &    $0.8$   &       -       &     $0.4326$   & $500\sim 600$     \\
H1705$-$250    &    $6-7$   &   $<1$    &     $5.2-8.6$  &    $0.3-0.6$ &    $0.8$   &       -       &     $0.5213$   & $500\sim 600$     \\
A0620$-$003    &  $\sim10$  &   $<1$    &    $11.0(19)$  &   $0.68(18)$ &    $0.6$   &       -       &     $0.3230$   &   $\sim 440$      \\
GS2000$+$251   &  $\sim10$  &   $<1$    &    $6.04-13.9$ &  $0.26-0.59$ &    $0.6$   &       -       &     $0.3441$   &   $\sim 440$      \\
\hline
\hline
\multicolumn{9}{|c|}{Nu: with evolved companion} \\
\hline
GRO J1655$-$40 &   $\sim5$  &  $1-2$    &    $5.1-5.7$   &   $1.1-1.8$  &    $0.8$   &  $0.65-0.75$  &   $2.6127(8)$  &   $\sim 430$      \\
4U 1543$-$47   &   $\sim5$  &  $1-2$    &    $2.0-9.7$   &   $1.3-2.6$  &    $0.8$   &  $0.75-0.85$  &     $1.1164$   &   $\sim 430$      \\
XTE J1550$-$564&  $\sim10$  &  $1-2$    &  $9.68-11.58$  & $0.96-1.64$  &    $0.5$   &       -       &    $1.552(10)$ &   $\sim 300$      \\
GS 2023$+$338  &  $\sim10$  &  $1-2$    &   $10.3-14.2$  & $0.57-0.92$  &    $0.5$   &       -       &     $6.4714$   &   $\sim 300$      \\
XTE J1819$-$254&    $6-7$   &  $\sim10$ &  $8.73-11.69$  & $5.50-8.13$  &    $0.2$   &               &      $2.817$   &  $10\sim 12$      \\
GRS 1915$+$105 &    $6-7$   &  $\sim10$ &     $14(4)$    &    $1.2(2)$  &    $0.2$   &    $>0.98$    &    $33.5(15)$  &  $10\sim 12$      \\
Cyg X$-$1      &    $6-7$   &$\gtrsim30$&    $\sim10.1$  &      $17.8$  &    $0.15$  &       -       &     $5.5996$   &    $5\sim6$       \\
\hline
\hline
\multicolumn{9}{|c|}{Extragalactic} \\
\hline
LMC X$-$1      &   $\sim40$ &  $\sim35$ &  $8.96-11.64$ & $30.62\pm3.22$& $\sim0.05$ &  $0.81-0.94$  &      $3.91$    &     $<2$          \\
LMC X$-$3      &     $7$    &    $4$    &     $5-11$    &    $6\pm2$    &    $0.43$  &    $>0.26$    &      $1.70$    &    $\sim155$      \\
M33 X$-$7      &   $\sim90$ &  $\sim80$ & $14.20- 17.10$&  $70.0\pm6.9$ & $\sim0.05$ &  $0.72-0.82$  &      $3.45$    &      $3-11$       \\
\hline
\end{tabular}
\caption{Parameters at the time of formation of BH and at present time.  Subindex $2$ stands for values at the time BH is formed, whereas subindex $now$ stands for recently measured values.  The AML binaries lose energy by GWs, shortening the orbital period whereas the Nu binaries will experience mass loss from the donor star to the higher mass BH and, therefore, move to longer orbital periods.}\label{tab-results}
\end{table}
}



\end{document}